\documentclass[aps,showpacs,superscriptaddress,twocolumn]{revtex4}%
\usepackage{amsfonts}
\usepackage{amsmath}
\usepackage{amssymb}
\usepackage{graphicx}%
\begin{document}
\title{Superfluidity and effective mass of magnetoexcitons in topological insulator
bilayers: Effect of inter-Landau-level Coulomb interaction}
\author{Zhigang Wang}
\affiliation{LCP, Institute of Applied Physics and Computational Mathematics, P.O. Box
8009, Beijing 100088, People's Republic of China}
\author{Zhen-Guo Fu}
\affiliation{State Key Laboratory for Superlattices and Microstructures, Institute of
Semiconductors, Chinese Academy of Sciences, P. O. Box 912, Beijing 100083,
People's Republic of China}
\affiliation{LCP, Institute of Applied Physics and Computational Mathematics, P.O. Box
8009, Beijing 100088, People's Republic of China}
\author{Ping Zhang}
\thanks{Corresponding author; zhang\_ping@iapcm.ac.cn}
\affiliation{LCP, Institute of Applied Physics and Computational Mathematics, P.O. Box
8009, Beijing 100088, People's Republic of China}
\thanks{To whom correspondence should be addressed. Email address: zhang\_ping@iapcm.ac.cn}


\begin{abstract}
The effective mass and superfluidity-normal phase transition temperature of
magnetoexcitons in topological insulator bilayers are theoretically
investigated. The intra-Landau-level Coulomb interaction is treated
perturbatively, from which the effective magnetoexciton mass is analytically
discussed. The inclusion of inter-Landau-level Coulomb interaction by more
exact numerical diagonalization of the Hamiltonian brings out important
modifications to magnetoexciton properties, which are specially characterized
by prominent reduction in the magnetoexciton effective mass and promotion in
the superfluidity-normal phase transition temperature at a wide range of
external parameters.

\end{abstract}
\maketitle

\section{Introduction}

The bilayer $n$-$p$ systems \cite{Lozovik, Shevchenko}, comprising electrons
from the $n$ layer and holes from the $p$ layer, have been the subject of
recent theoretical and experimental investigations. These systems, including
coupled quantum wells \cite{Snoke,Butov, Timofeev, Eisenstein}\ and layered
graphene \cite{Zhang,Min,Lozovik1,Iyengar,Berman, Berman1,Koinov}, are of
interest, in particular, in connection with the possibility of the
Bose-Einstein condensation and superfluidity of indirect excitons or
electron-hole pairs. The superfluidity is manifested as nondissipative flow
of electric currents, equal in magnitude and opposite in direction, along
the layers. In high magnetic fields, two-dimensional excitons survive in a
substantially wider temperature range, as the exciton binding energies
increase with magnetic field.

Recently, topological insulator (TI) as a new phase of quantum matter,
which can not be adiabatically connected to conventional insulators and
semiconductors, have been studied intensively \cite%
{Kane2005,Bernevig,Fu2007,Konig,Hsieh,Zhang2009,Xia2009,Chen2009,Hasan,Qi2}.
Present technological advances have allowed the production of topological insulator bilayers (TIBs) system, which
consist of two TI thin films separated by a dielectric
barrier.
On one hand, CdTe/HgTe TI quantum wells are two-dimensional which are ideal for designing the TIBs, since
CdTe/HgTe can be either doped in \textit{n}-type or \textit{p}-type \cite{LuoJW}; while on the other hand, although the three-dimensional
TI Bi$_{2}$Se$_{3}$ is dominated by charged selenium vacancies, which
results in \textit{n}-type behavior, it could also exhibit \textit{p}-type behavior by doping method.
Transport and magnetic properties of TIs doped by Cr, Fe, and Cu, and \textit{n}-type to
 \textit{p}-type crossover in  Bi$_{2}$Se$_{3}$ codoped with Sb and Pb have been investigated
 variously in experiments.
Recently, in particular, the \textit{p}-type Bi$_{2}$Se$_{3}$ single crystal has been obtained by doping Ca in experiments \cite{YSHor}.
 It is possible to use Al$_{2}$O$_{3}$ or SiO$_{2}$ as the
dielectric spacer, on which, remarkably, high-quality Bi$_{2}$Se$_{3}$ and Bi%
$_{2}$Te$_{3}$ TI quantum well thin films have been now successively grown
\cite{Chang,Li,Liu,Aguilar}. Besides, high quality Bi$_{2}$Se$_{3}$/ZnSe multilayers \cite{XieMH2011},
and superlattices constructed by alternating Bi$_{2}$Se$_{3}$ and In$_{2}$Se$_{3}$ layers \cite{XieMH20112}
have also been successfully obtained in experiments.

TIs are characterized by a full insulating gap in the bulk and protected
gapless edge or surface states. It is the gapless edge or surface states
that results in the absence of exciton at the TI heterostructures. However, there are
various methods, such as introduction
of a magnetic exchange field or a strong external magnetic field, to produce
 a gap in TI heterostructures, which may lead to the formation of excitons \cite{Franz1,Hao} or
magnetoexcitons. Magnetoexcitons, which have been observed experimentally in semiconductor quantum wells, is
 an ideal object in exploring the
Coulomb interaction effects in TIBs system and Bose-Einstein condensation
since it behaves as neutral bosons at low densities. Moreover, as mentioned above,
the TIBs, which are required for generating
magnetoexcitons, are achievable in current experimental capabilities.
Motivated by its importance both from basic point of interest and from application of TI-based electronics,
in the present paper we address this issue by presenting an
attempt at the theoretical evaluation of superfluidity property and effective mass of magnetoexcitons in the
TIBs structure [see Fig. 1(a)] in the presence of a perpendicular high magnetic field, which produces a gap since the Dirac-type
energy spectrum becomes discrete by forming Landau levels (LLs).
\begin{figure}
\begin{center}
\includegraphics[width=1.\linewidth]{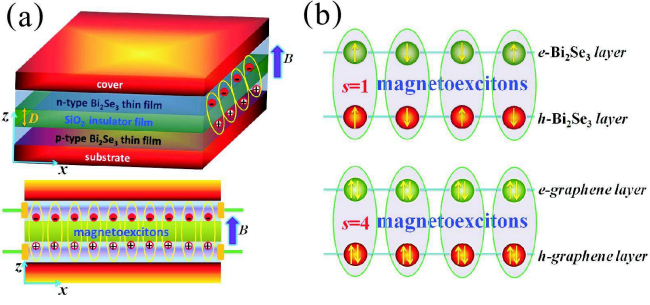}
\end{center}
\caption{(Color online) (a) Scheme of TIB comprising two Bi$_{2}$Se$_{3}$
thin films separated by SiO$_{2}$ spacer. Electron (hole) carriers are
induced by $n$ ($p$)-type doping or applied gates. To produce
magnetoexcitons a strong perpendicular magnetic field is applied. (b)
Illustration of the difference of magnetoexcitons between TIB and graphene
bilayer. The spin degeneracy factor $s$=$4$ in graphene bilayer leads its
undressed magnetoexciton mass to be $1/4$ times of that in TIB.}
\end{figure}

In the presence of the high magnetic
field, the system may exhibit the phase transition from the
high-temperature disordered or delocalized phase with the exponential
correlation to a low-temperature quasi-ordered or localized phase, which is
named as Kosterlitz-Thouless (KT) transition. We study
the properties of magnetoexcitons in TIBs, including the effective
magnetoexciton mass $m_{B}$ and superfluid-normal state, i.e.,
KT transition temperature $T_{c}$. Comparing to
bilayer graphene \cite{Berman}, the effective mass of magnetoexcitons in
TIBs is found to be four times larger due to the spin degeneration [see Fig. %
1(b)]. Meanwhile, we also investigate the effect of inter-LL
Coulomb interaction on the magnetoexciton superfluidity in TIBs, which has
been ignored in previous studies on semiconductor and graphene systems. We
find that although the intra-LL Coulomb interaction plays the main role in
deciding the effective magnetoexciton mass and KT temperature, the inclusion
of inter-LL Coulomb interaction brings out important modifications. In fact,
at the magnetic field $B$=$10\sim 20$ T and the spacer thickness $D$=$25\sim
35$ nm in the setup in Fig. 1(a), our results show that the inter-LL Coulomb
interaction will reduce the magnetoexciton effective mass by 15\%$\sim $25\%
and improve the KT temperature by 20\%$\sim $25\%.

\section{Effective magnetic mass}

We start with the effective Hamiltonian of TIB system $H=H_{0}+U(r)$, where
\begin{equation}
H_{0}=v_{F}\mathbf{\sigma }_{e}\cdot \left( \mathbf{\hat{z}}\times \mathbf{%
\pi }_{e}\right) -v_{F}\mathbf{\sigma }_{h}\cdot \left( \mathbf{\hat{z}}%
\times \mathbf{\pi }_{h}\right)  \label{Hami}
\end{equation}%
is the free electron-hole part of TIB, while $U(r)=-e^{2}/\epsilon \sqrt{|%
\mathbf{r}|^{2}+D^{2}}$ is the Coulomb interaction between a pair of
electron and hole with $r=r_{e}-r_{h}$ being the relative coordinate in the $%
x-y$ plane and $D$ the spacer thickness. Here, $r_{e(h)}$ represents the
position of the electron (hole), and $\pi _{e(h)}$ $=p_{e(h)}\pm
eA_{e(h)}/c=-i\partial /\partial r_{e(h)}\pm eA_{e(h)}/c$ denotes the
in-plane momentum of the electron (hole), where the gauge is chosen as $%
A_{e(h)}=B(0,x_{e(h)},0)$ in the following calculations. $v_{F}$ is the
Fermi velocity ($\sim 3\times 10^{5}$ m/s for Bi$_{2}$Se$_{3}$-family
materials), $\hat{z}$ is the unit vector normal to the surface, and $\sigma
_{e}=$ $\sigma \otimes I$ ($\sigma _{h}=$ $I$ $\otimes $ $\sigma $)
describes the spin operator acting on the electron (hole), in which $\sigma $
denotes a vector of Pauli matrices and $I$ is the 2$\times $2 identity
matrix.

The LLs of $H_{0}$ are given by
\begin{equation}
E_{n_{+},n_{-}}^{(0)}=\frac{\sqrt{2}v_{F}}{r_{B}}[\text{sgn}(n_{+})\sqrt{%
|n_{+}|}-\text{sgn}(n_{-})\sqrt{|n_{-}|}].  \label{f14}
\end{equation}%
The corresponding eigenstates can be analytically expressed as \cite{Iyengar}
\begin{equation}
\psi _{\mathbf{P}}\left( \mathbf{R},\mathbf{r}\right) =\exp \left[ i\left(
\mathbf{P}+\frac{e}{2c}\left[ \mathbf{B}\times \mathbf{r}\right] \right)
\cdot \mathbf{R}\right] \Psi \left( \mathbf{r}-\mathbf{r}_{0}\right) ,
\label{f3}
\end{equation}%
where $R=(r_{e}+r_{h})/2$ and $r_{0}=r_{B}^{2}(\hat{B}\times P)$ with
magnetic length $r_{B}=\sqrt{\hbar c/eB}$ and unit magnetic field $\hat{B}$.
For an electron in LL $n_{+}$ and a hole in LL $n_{-}$, the four-component
wave functions in the relative coordinate are written as
\begin{align}
& \Psi _{n_{+},n_{-}}(\mathbf{r})=|n_{+},n_{-}\rangle  \label{f13} \\
& =(\sqrt{2})^{\delta _{n_{+},0}+\delta _{n_{-},0}-2}\left(
\begin{array}{c}
\Phi _{|n_{+}|-1,|n_{-}|-1}(\mathbf{r}) \\
i^{-sgn(n_{-})}\Phi _{|n_{+}|-1,|n_{-}|}(\mathbf{r}) \\
i^{sgn(n_{+})}\Phi _{|n_{+}|,|n_{-}|-1}(\mathbf{r}) \\
i^{sgn(n_{+})-sgn(n_{-})}\Phi _{|n_{+}|,|n_{-}|}(\mathbf{r})%
\end{array}%
\right) ,  \notag
\end{align}%
where%
\begin{align}
\Phi _{n_{1},n_{2}}(\mathbf{r})& =\frac{2^{-\frac{|l_{z}|}{2}}}{\sqrt{2\pi }}%
\frac{n_{\_}!}{\sqrt{n_{1}!n_{2}!}}\frac{1}{r_{B}}  \label{f5} \\
& \times e^{-il_{z}\phi }sgn(l_{z})^{l_{z}}\frac{r^{|l_{z}|}}{r_{B}^{|l_{z}|}%
}L_{n_{-}}^{|l_{z}|}(r^{2}/2r_{B}^{2})e^{-r^{2}/4r_{B}^{2}}  \notag
\end{align}%
with $L$ denoting Laguerre polynomials $z=x+iy$, $l_{z}=n_{1}-n_{2}$, $%
n_{-}=\min (n_{1},n_{2})$, $z/|z|=e^{i\phi }$, and sgn$(l_{z})^{l_{z}}%
\rightarrow 1$ for $l_{z}=0$.\begin{figure}
\begin{center}
\includegraphics[width=1.\linewidth]{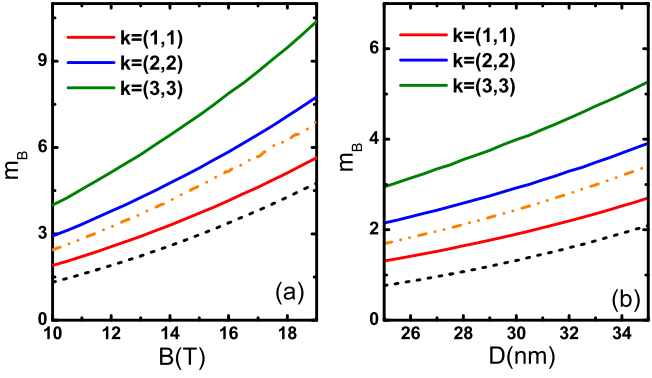}
\end{center}
\caption{(Color online) Calculated magnetoexciton effective mass $m_{B}$
(in unit of free electron mass) as functions of (a) magnetic field $B$ and
(b) spacer width $D$, for different choices of magnetoexciton LL indices $k$=%
$(n,n)$. For comparison, the $k$=$(1,1)$ magnetoexciton effective mass
without including inter-LL transition (given by Eq. (\protect\ref{f13})) is
plotted by dash-dot-dotted lines and $m_{B}^{(0)}(D)$ given by Eq. (\protect
\ref{f25}) is shown by dashed lines. The spacer width in panel (a) is set as
$D$=$30$ nm and the external magnetic field in panel (b) is set as $B$=$10$
T in (b). The other parameters are set as $v_{F}$=$3.0\mathtt{\times }10^{5}$
m/s and $\protect\epsilon $=$4.0$.}
\end{figure}

When the inter-LL Coulomb interaction is taken into account, however, one
should perform diagonalization of full Hamiltonian of Coulomb interacting
carriers in some basis of magnetoexcitonic states $\Psi _{n_{+},n_{-}}(r)$.
In other words, to obtain eigenvalues of the Hamiltonian $H$, we need to
numerically solve the following eigen-equation \cite{Lozovik2011}:

\begin{align}
0& \mathtt{=}\det \left\Vert \delta _{n_{+},n_{+}^{\prime }}\delta
_{n_{-},n_{-}^{\prime }}(E_{n_{+},n_{-}}^{(0)}\mathtt{-}E)\right.
\label{f16} \\
& \left. \mathtt{+}\langle \Psi _{n_{+}^{\prime },n_{-}^{\prime }}|U\left(
\mathbf{r}\mathtt{-}r_{B}^{2}\mathbf{\hat{z}}\mathtt{\times }\mathbf{P}%
\right) |\Psi _{n_{+},n_{-}}\rangle \right\Vert .  \notag
\end{align}%
The location of the chemical potential will determine the possible LL
indices for electrons and holes. We should point out that the intra-LL component of the
Coulomb interaction is defined as $\langle\Psi_{n_{+},n_{-}}|U|\Psi_{n_{+}%
,n_{-}}\rangle$, while the inter-LL component is defined as $\langle
\Psi_{n_{+}^{\prime},n_{-}^{\prime}}|U|\Psi_{n_{+},n_{-}}\rangle$, where
$|\Psi_{n_{+}^{\prime},n_{-}^{\prime}}\rangle\neq|\Psi_{n_{+},n_{-}}\rangle$.

Consider a magnetoexciton is formed by an electron on the LL $n$ and a hole
on the LL $m$. In the limit of the relatively large separation $D$ between
electron and hole TIBs and relatively high magnetic field $B$ when $%
e^{2}/\left( \epsilon D\right) \ll \hslash v_{F}/r_{B}$, the magnetoexciton
energy can be approximated by only considering its zeroth order energy part $%
E_{n,m}^{(0)}$. However, at higher magnetic field $10\sim 20$ T in
experiment, the Coulomb interaction $e^{2}/\left( \epsilon D\right) $ is
only several times less than the zeroth energy $\hslash v_{F}/r_{B}$. In
this case, the Coulomb interaction can be treated as a perturbation, which
could be departed into intra-LL part and inter-LL part. For the intra-LL
part, the magnetoexciton energy can be written as%
\begin{align}
E_{n,m}& =E_{n,m}^{(0)}+\langle \Psi _{n,m}|U\left( \mathbf{r}-r_{B}^{2}%
\mathbf{\hat{z}}\times \mathbf{P}\right) |\Psi _{n,m}\rangle   \notag \\
& =E_{n,m}^{(0)}+2^{\delta _{n,0}+\delta _{m,0}-2}  \notag \\
& \times \left\{ \langle \langle |n|-1,|m|-1,\mathbf{P}|U||n|-1,|m|-1,%
\mathbf{P}\rangle \rangle \right.   \notag \\
& +\langle \langle |n|-1,|m|,\mathbf{P}|U||n|-1,|m|,\mathbf{P}\rangle
\rangle   \notag \\
& +\langle \langle |n|,|m|-1,\mathbf{P}|U||n|,|m|-1,\mathbf{P}\rangle
\rangle   \notag \\
& \left. +\langle \langle |n|,|m|,\mathbf{P}|U||n|,|m|,\mathbf{P}\rangle
\rangle \right\} ,  \label{f77}
\end{align}%
where the notation $\langle \langle nmP|U|nmP\rangle \rangle $ denotes the
averaging by the two-dimensional harmonic oscillator eigenfunctions $\Phi
_{n,m}(r)$. Substituting for small magnetic momenta $P\ll \hslash /r_{B}$
and $P\ll \hslash D/r_{B}^{2}$ the relation%
\begin{equation}
\langle \langle nmP|U|nmP\rangle \rangle =\mathcal{E}_{nm}^{(b)}+\frac{P^{2}%
}{2M_{nm}(B,D)}  \label{f10}
\end{equation}%
into Eq. (\ref{f77}), we can get the dispersion law of a magnetoexciton for
small magnetic momenta. As an example, let us consider the magnetoexciton on
the $k$=$\left( 1,1\right) $ LL. The magnetoexciton energy at small magnetic
momenta is then given by
\begin{equation}
E_{1,1}(P)=\mathcal{E}_{B}^{(b)}(D)+\frac{P^{2}}{2m_{B}(D)},  \label{f11}
\end{equation}%
where the binding energy $E_{B}^{(b)}(D)$ is expressed as
\begin{equation}
\mathcal{E}_{B}^{(b)}(D)=\frac{1}{4}E_{00}^{(b)}+\frac{1}{2}E_{01}^{(b)}+%
\frac{1}{4}\mathcal{E}_{11}^{(b)}
\end{equation}%
and the effective magnetic mass $m_{B}(D)$ of a magnetoexciton is expressed
as%
\begin{equation}
\frac{1}{m_{B}}=\frac{1}{4M_{00}}+\frac{1}{2M_{01}}+\frac{1}{4M_{11}}.
\label{f13}
\end{equation}%
Here, the constants $E_{00}^{(b)}$, $E_{01}^{(b)}$, $E_{11}^{(b)}$, $M_{00}$%
, $M_{01}$, and $M_{11}$ depend on magnetic field $B$ and the interlayer
separation $D$. Explicitly, $E_{00}^{(b)}=-CE_{0}$, $E_{01}^{(b)}=-E_{0}[(%
\frac{1}{2}-x)C+\frac{x}{\sqrt{\pi }}]$, $E_{11}^{(b)}=-E_{0}[(\frac{3}{4}%
+x^{2}+x^{4})C-\frac{1}{\sqrt{\pi }}(\frac{x}{2}+x^{3})]$, and $%
M_{00}=M_{0}[\left( 1+2x^{2}\right) C-\frac{2x}{\sqrt{\pi }}]^{-1}$, $%
M_{01}=M_{0}[\left( 3+2x^{2}\right) \frac{x}{\sqrt{\pi }}-(\frac{1}{2}%
+4x^{2}+2x^{4})C]^{-1}$, $M_{11}=M_{0}[\frac{C}{4}\left(
7+50x^{2}+44x^{4}+8x^{6}\right) -(\frac{17}{2}+10x^{2}+2x^{4})\frac{x}{\sqrt{%
\pi }}]^{-1}$, with $x=D/(\sqrt{2}r_{B})$, $E_{0}=\frac{e^{2}}{\epsilon r_{B}%
}\sqrt{\frac{\pi }{2}}$, $M_{0}=\frac{2^{3/2}\epsilon \hslash ^{2}}{\sqrt{%
\pi }e^{2}r_{B}}$, and functions $erfc\left( x\right) $ and $C(x)$
respectively defined as $erfc(x)=\frac{2}{\sqrt{\pi }}%
\int_{x}^{+\infty }\exp (-t^{2})dt$ and $C\left( x\right) =e^{x^{2}}erfc\left( x\right) $. In the special limit of $D\gg r_{B}$, one obtains
\begin{equation}
m_{B}^{(0)}(D)=\frac{D^{3}\hslash ^{2}\epsilon }{r_{B}^{4}e^{2}}.
\label{f25}
\end{equation}%
We should notice that the effective magnetic mass in the present system is
equal to that in coupled semiconductor quantum wells at the same $D$, $%
\epsilon $, and $B$, which is four times larger than that in graphene
bilayers \cite{Berman, Koinov}. Although the effective low-energy
Hamiltonian of the electron-hole graphene bilayer system is also expressed
as a $4\times 4$ matrix, which is similar to Eq. (\ref{Hami}) shown above,
the momentum therein is actually coupled with the pseudospins of sublattices
rather than the real spins of electron and hole. Therefore, the factor $1/4$
in the effective magnetic mass of magnetoexcitons in graphene bilayer system
arises from the spin degeneracy (see Fig. 1(b)), which is absent
in the TIB system considered here since the freedom of spin has been
involved due to the strong spin-orbit coupling in the original effective
Hamiltonian Eq. (\ref{Hami}). This significant difference between TIB and
graphene bilayer may be detected by the gyromagnetic resonance experiments.

Now let us make a simple estimate to illustrate the reasonability of the
perturbation theory. We choose the Fermi velocity and the electron-hole
interlayer distance as $v_{F}=3\times 10^{5}$ m/s and $D=30$ nm. The
dielectric constant of the spacer SiO$_{2}$ is $\epsilon =4.0$. Under these
conditions, one can obtain the Coulomb interaction $e^{2}/\left( \epsilon
D\right) =12$ meV and kinetic energy $\hslash v_{F}/r_{B}=24.4$ meV
(corresponding to $r_{B}=8.1$ nm) when the applied magnetic field $B=10$ T.
In this case, $e^{2}/\left( \epsilon D\right) /\left( \hslash
v_{F}/r_{B}\right) \simeq 0.5$, which indicates that in our studied TIB
system, the effect of Coulomb interaction can be treated as a perturbation.
Same analysis can also be applied to graphene, in which the Fermi velocity
is about $3$ times larger than that in Bi$_{2}$Se$_{3}$ film. With the same
external parameters for estimation, one can immediately obtain $e^{2}/\left(
\epsilon D\right) /\left( \hslash v_{F}/r_{B}\right) \simeq 0.17$ for
graphene bilayer system. In the above calculations of the magnetoexciton
energy and effective mass we only consider the intra-LL Coulomb interaction.
On the other hand, the inter-LL Coulomb interaction is also important. We
will show in the following that the inter-LL Coulomb interaction can bring
about 0.15$\sim $0.25 times modification of the magnetoexciton effective
mass (see Fig. 2).\begin{figure}
\begin{center}
\includegraphics[width=1.\linewidth]{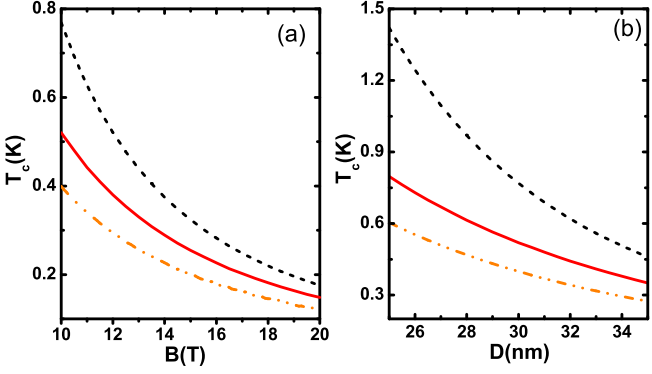}
\end{center}
\caption{(Color online) Calculated KT temperature versus (a) magnetic
field and (b) spacer width. The density of magnetoexciton is set as $n$=4.0$%
\times 10^{11}$ cm$^{-2}$. The spacer width $D$=$30$ nm in used in panel (a)
and the magnetic field $B$=$10$ T is used in panel (b). The red solid, black dashed,
and blue dotted lines correspond to the magnetoexciton effective mass
obtained by considering both intra-LL and inter-LL Coulomb interaction, only
intra-LL Coulomb interaction, and the approximation Eq. (\protect\ref{f25}%
) at $D\mathtt{\gg }r_{B}$, respectively.}
\end{figure}

To study the whole effect of Coulomb interaction, including the intra-LL and
inter-LL Coulomb interaction, on the effective magnetoexciton mass as well
as KT transition to the superfluid state, we have to numerically solve Eq. (%
\ref{f16}). In our numerical calculations we have used four electron and
four hole levels around the LL $k=\left( n,n\right) $. Figures 2%
(a) and 2(b) respectively show the numerically calculated
magnetoexciton mass $m_{B}(D)$ as a function of the magnetic field and the
dielectric spacer width $D$ with different LL indices $k$. To observe the
effect of inter-LL Coulomb interaction, we also exhibit in both figures the
magnetoexciton effective mass at LL $k=\left( 1,1\right) $ by dash-dotted
lines without inter-LL transition [Eq. (\ref{f13})], and the analytical
approximation results of magnetoexciton mass $m_{B}^{(0)}$ [Eq. (\ref{f25})]
by dashed lines. Comparing these results, one can find that for the magnetic
field strength $B=10\sim 20$T and the spacer width $D=25\sim 35$ nm, the
magnetoexciton effective mass is reduced by $15\%\sim 25\%$ by the inclusion
of the inter-LL Coulomb interaction.

\section{Superfluidity of dipole magnetoexcitons}

The magnetoexcitons that are constructed by spatially separated electrons
and holes in TIB at large interlayer separation $D\gg r_{B}$ form
two-dimensional weakly nonideal gas of bosons. This is valid, because we can
think the indirect exciton interact as parallel dipoles when $D\gg r_{B}$.
So at low temperatures, these magnetoexcitons condense in Bose-Einstein type
and form superfluid state below the KT transition temperature $T_{c}$ \cite%
{Thouless, Kosterlitz}, which is determined by%
\begin{equation}
T_{c}=\frac{\pi \hslash ^{2}n_{s}\left( T_{c}\right) }{2k_{B}m_{B}},
\label{f30}
\end{equation}%
where $n_{s}\left( T\right) $ is the superfluid density of the
magnetoexciton system and $k_{B}$ is the Boltzmann constant.

To solve the superfluid density $n_{s}$ in Eq. (\ref{f30}), we can first
solve the normal density $n_{n}$ because $n_{s}=n-n_{n}$, where $n$ is the
total density. Through the usual procedure \cite{Abrikosov, Berman}, we have
the normal density $n_{n}$ as
\begin{equation}
n_{n}=\frac{3\zeta (3)}{2\pi \hslash ^{2}}\frac{k_{B}^{3}T^{3}}{%
m_{B}c_{s}^{4}},  \label{f31}
\end{equation}%
where $\zeta (z)$ is the Riemann zeta function and $c_{s}=\sqrt{\mu /m_{B}}$
is the sound velocity with the chemical potential
\begin{equation}
\mu =\frac{\pi \hslash ^{2}n}{m_{B}\ln \left[ \hslash ^{4}\epsilon
^{2}/(2\pi nm_{B}^{2}e^{4}D^{4})\right] }.  \label{f32}
\end{equation}%
Finally, combining Eqs. (\ref{f30})-(\ref{f32}), we obtain the KT transition
temperature%
\begin{align}
T_{c}& =\left[ \left( 1+\sqrt{\frac{32}{27}\left( \frac{m_{B}k_{B}T_{c}^{0}}{%
\pi \hslash ^{2}n}\right) ^{3}+1}\right) ^{1/3}\right.   \label{f33} \\
& \left. -\left( \sqrt{\frac{32}{27}\left( \frac{m_{B}k_{B}T_{c}^{0}}{\pi
\hslash ^{2}n}\right) ^{3}+1}-1\right) ^{1/3}\right] \frac{T_{c}^{0}}{2^{1/3}%
},  \notag
\end{align}%
where $T_{c}^{0}=\frac{1}{k_{B}}\left( \frac{2\pi \hslash ^{2}n\mu ^{2}}{%
3m_{B}\zeta (3)}\right) ^{1/3}$is an auxiliary quantity that is equal to the
temperature at which the superfluid density vanishes in the mean-field
approximation (i.e., $n_{s}(T_{c}^{0})=0$).
\begin{figure}
\begin{center}
\includegraphics[width=1.\linewidth]{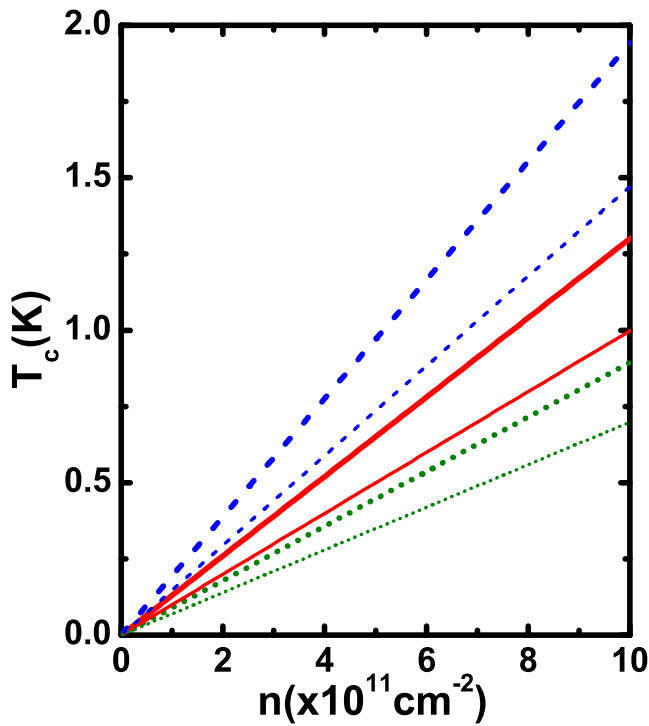}
\end{center}
\caption{(Color online) Calculated KT temperature versus magnetoexciton
density $n$ at $B$=$10$ T. The solid, dashed, and dotted curves correspond
to $D$=$30$, $25$, and $35$ nm, respectively. Here for comparison, the
results with and without inter-LL Coulomb interaction included are given by
thick and thin curves, respectively.}
\end{figure}

We calculate $T_{c}$ for different choices of magnetic field ($10\sim 20$
T), dielectric spacer width $D$ ($25\sim 35$ nm), and magnetoexciton density
$n$ (up to $1.0\times 10^{12}$ cm$^{-2}$). The results are presented in Fig. %
3 and Fig. 4. One can find the following features:
(i) The KT temperature is up to several Kelvins in our calculated ranges,
which is similar to that in graphene bilayer and coupled semiconductor
quantum wells \cite{Berman,Berman1}; (ii) Figure 3 clearly shows
that the calculated $T_{c}$ evidently depends on the effective
magnetoexciton mass form. Comparing the $T_{c}$ calculated with full Coulomb
interaction (red lines) with that in the absence of inter-LL Coulomb
interaction (orange lines), one can find that there is about 20\%$\sim $25\%
correction induced by inter-LL Coulomb interaction. By decreasing the
magnetic field or the dielectric spacer width, this correction becomes more
prominent, which indicates that the inter-LL transition caused by Coulomb
interaction is significant in estimating $T_{c}$. When the magnetoexciton
mass approximately takes the limit form Eq. (\ref{f25}) at $D\gg r_{B}$, the
KT temperature has a promotion (black lines). Especially at relatively small
values of $B$ and $D$, this promotion becomes very remarkable and almost
reaches $100\%$; (iii) Figure 4 shows $T_{c}$ in good
approximation linearly increases with increasing magnetoexciton density $n$.
This is due to the fact that the denominator in Eq. (\ref{f32}) for the
chemical potential and thus the sound velocity, weakly depends on $n$.

\section{Conclusion}

In summary, we have theoretically studied the effective mass and KT
transition temperature of magnetoexcitons in TIB structure under a strong
perpendicular magnetic field. When only intra-LL Coulomb interaction is
considered, the effective magnetoexciton mass in TIB structure is four times
larger than that in graphene bilayer structure, while the calculated KT
temperature is about several Kelvins, same in amplitude as in graphene
bilayer and semiconductor bilayer systems. The inclusion of
inter-Landau-level Coulomb interaction has been shown to bring about
significant corrections to magnetoexciton properties by prominently reducing the magnetoexciton
effective mass and promoting the KT temperature.

\begin{acknowledgments}
This work was supported by NSFC under Grants No. 90921003 and No. 10904005,
and by the National Basic Research Program of China (973 Program) under Grant
No. 2009CB929103.
\end{acknowledgments}

\end{document}